\documentstyle[12pt]{article}

\newcommand{\gsim}{\buildrel > \over {_\sim}}

\newcommand{\pabs}{{|\bf p|}}
\newcommand{\kta}{{\bf k}_{\perp a}}
\newcommand{\ktb}{{\bf k}_{\perp b}}
\newcommand{\QT}{{\bf Q}_\perp}
\newcommand{\ie}{{\it i.e.}}
\newcommand{\eg}{{\it e.g.}}
\newcommand{\cf}{{\it c.f.}}
\newcommand{\order}{{\cal O}}
\newcommand{\half}{{1\over 2}}
\newcommand{\gev}{{\rm \ GeV}}

\newcommand{\PLB}[3]{\mbox{}Phys. Lett. {\bf B{#1}}, {#2} ({#3})}
\newcommand{\NPB}[3]{\mbox{}Nucl. Phys. {\bf B{#1}}, {#2} ({#3})}
\newcommand{\PRL}[3]{\mbox{}Phys. Rev. Lett. {\bf {#1}}, {#2} ({#3})}
\newcommand{\PRD}[3]{\mbox{}Phys. Rev. {\bf D{#1}}, {#2} ({#3})}
\newcommand{\ZPC}[3]{\mbox{}Z. Phys. {\bf C{#1}}, {#2} ({#3})}

\newcommand{\etal}{{\em et al.}}

\begin{document}
\setlength{\baselineskip}{7mm}

\titlepage
\begin{flushright}
HU-TFT-94-12\\
LBL-35430\\
hep-ph/9404322
\end{flushright}
\vskip2.0truecm
\begin{center}
{\large {\bf Higher-Twist Effects in the Drell-Yan Angular Distribution}}
\\[3ex]
\vskip1.0truecm
\setcounter{footnote}{1}
{\large {K.J. Eskola$^1$, P. Hoyer$^1$, M. V\"anttinen$^2$, R.
Vogt$^3$}}\\[3ex]
\footnotetext{
 Laboratory of High Energy Physics, P.O. Box 9,
FIN-00014, University of Helsinki, Finland.

$^2$ Research Institute for Theoretical Physics, P.O. Box 9,
FIN-00014, University of Helsinki, Finland.

$^3$ Nuclear Science Division, Lawrence Berkeley Laboratory,
Berkeley, CA 94720, USA. Supported by the U. S. Department of Energy under
Contract No. DE-AC03-76SF0098.

}
\setcounter{footnote}{0}
\end{center}
\begin{abstract}

We study the Drell-Yan process $\pi N \rightarrow \mu^+ \mu^- X$ at
large $x_F$ using perturbative QCD.
A higher-twist mechanism suggested by Berger and Brodsky
is known to qualitatively explain the observed
$x_F$ dependence of the muon angular distribution, but
the predicted large $x_F$ behavior differs
quantitatively from observations.
We have repeated the model calculation taking into account the effects of
nonasymptotic
kinematics.  At fixed-target energies we find important
corrections which improve the agreement with data. The asymptotic
result of Berger and Brodsky is recovered only at much higher energies.  We
discuss the generic reasons for the large corrections at high $x_F$. A proper
understanding of the $x_F \to 1$ data would give important information on the
pion distribution amplitude and exclusive form factor.

\end{abstract}
\clearpage

The Drell-Yan process of inclusive muon pair production in hadronic collisions
\cite{DrellYan} is the most important means of determining the valence parton
content of hadrons other than the nucleon.
The kinematics are
defined by the longitudinal-momentum fraction of the muon pair, $x_F$, the
squared invariant mass of the pair, $Q^2$, and the transverse momentum,
$\QT$,  of
the pair. In the Drell-Yan model the reaction is described as the annihilation
of a massless quark and an antiquark into a virtual photon which decays into
a muon pair. Because of rotational invariance and parity conservation, the
polar angle distribution of massless muons in the muon pair rest frame
must be of the form
\begin{equation}
  1 + \lambda \cos^2 \theta. \label{deflambda}
\end{equation}
In the Drell-Yan model $\lambda = 1$ corresponding to the fact that the
annihilation of on-shell spin-$\half$ particles produces a transversely
polarized virtual photon. This expectation is maintained within small
corrections in the presence of order $\alpha_s(Q^2)$
perturbative QCD radiative corrections
\cite{Chiappetta}.

Experimentally, a transverse photon polarization is observed
in most of the kinematic region. At large $x_F$, however, the photon appears
to become longitudinally polarized \cite{NA10,Conway}.  In the following, we
shall study a higher-twist mechanism leading to longitudinal polarization,
within a model originally suggested by Berger and Brodsky (BB)
\cite{BergerBrodsky}.

To produce a large $x_F$
muon pair, it is necessary that $x_a$, the momentum
fraction carried by the annihilating
projectile parton, is close to 1.
Fock states $F$ of the projectile $H$
where one parton carries most of the momentum have large energy,
\begin{equation}
E_H-E_F = \Delta E \simeq - {\kta^2\over 2p(1-x_a)} \label{ediff}
\end{equation}
where $-\kta$ and $(1-x_a)p$ are the transverse and longitudinal momenta of the
soft parton(s) in the Fock state. In the limit $x_a \to 1$, the lifetime of
the Fock state, $\tau_F = 1/\Delta E \propto 1-x_a$ is short, and perturbation
theory is applicable. In the BB model, the leading contribution with
a pion projectile is then
given by the diagrams of Fig. 1, where the spectator valence quark transfers
its
momentum to the active quark via single gluon exchange.
Thus the whole pion state participates in the scattering,
implying longitudinal polarization of the virtual photon due to
helicity conservation in the pion-virtual photon interaction.

As a first approximation, BB took the nonperturbative wave function of the pion
to be $\delta(z-\half)$, \ie, both valence quarks carry
half of the pion
momentum\footnote{This assumption was relaxed and the BB model generalized
in Ref.\cite{DYworkshop}}. In the limit
\begin{equation}
 x_a \rightarrow 1, \;\;\; \kta^2 / Q^2 \rightarrow 0\ \  {\rm with}\ \
\kta^2 / Q^2  \sim (1-x_a)^2,
 \label{BBlimit}
\end{equation}
the differential cross section is proportional to \cite{BergerBrodsky}
\begin{equation}
 (1-x_a)^2 (1 + \cos^2 \theta) + \frac{4{\bf k}_{\perp a}^2}{9Q^2} \sin^2
 \theta. \label{BBexpr}
\end{equation}
This gives a simple expression for the parameter $\lambda$ in the
angular distribution (\ref{deflambda}),
\begin{equation}
\lambda = {(1-x_a)^2-4\kta^2/9 Q^2\over (1-x_a)^2+4\kta^2/9 Q^2}.
\label{BBlamb}
\end{equation}
It is clear that if $(1-x_a)^2 $
vanishes faster than  $\kta^2 / Q^2$, the angular distribution turns into
$ \sin^2 \theta $, and the parameter $\lambda$ approaches $-1$.
Experimentally, a drop in $\lambda$ at large $x_F$ is observed \cite{Conway},
but the data lie well above the curve derived from the
asymptotic expression (\ref{BBexpr}).  The purpose of this letter is to show
that there are important nonasymptotic corrections to the limiting expression
(\ref{BBexpr}) at present energies. The BB model evaluated with exact
kinematics is, in fact, in rather good agreement with the data.

The muon pair production cross section resulting from the model
of Fig. 1, without taking the limit (\ref{BBlimit}) and after integrating
over the azimuthal angle of the $\mu^+$, is
\begin{eqnarray}
 \frac{Q^2 d\sigma^{\pi^-N\rightarrow \mu^+\mu^- X}}{dQ^2 d^2 \QT dx_F
d\cos\theta}  \propto
 (\alpha \alpha_s \psi_\pi({\bf r}_\perp =0))^2 |{\bf p}|
 \int \frac{dx_a dx_b d^2 \ktb d^2 \kta} {\sqrt{\kta^2 + (1-x_a)^2 {\bf p}^2}}
 \nonumber \\
\times  \left[e_u^2f_{u/N}(x_b,\ktb) + e_d^2f_{\bar d/N}(x_b,\ktb)
\right]\frac{1}{s t^2} \left[ \varrho_{11} (1+\cos^2 \theta) + \varrho_{00}
\sin^2 \theta \right]
  \nonumber \\
\times\delta(Q^2 + \QT^2 + x_F^2 {\bf p}^2
 - \left[ \pabs + \sqrt{\ktb^2 + x_b^2 {\bf p}^2}
 - \sqrt{\kta^2 + (1-x_a)^2 {\bf p}^2} \right] ^2)
  \nonumber \\
\times\delta(x_F-x_a+x_b) \delta(\QT-\kta-\ktb)
  \label{exact}
\end{eqnarray}
where ${\bf p}$ is the pion momentum in the hadron center-of-mass frame,
and we have expressed the
invariants at the $\pi$-parton level as
\begin{eqnarray}
s & = & 2 p \cdot p_b
         = 2|{\bf p}|(\sqrt{\ktb^2 + x_b^2 {\bf p}^2} + x_b|{\bf p}|) \\
t & = & -2 p \cdot p_1
  = -2 |{\bf p}|(\sqrt{\kta^2 + (1-x_a)^2 {\bf p}^2} - (1-x_a)|{\bf p}|) \,
\, ,
\label{t}
\end{eqnarray}
neglecting all masses but $Q^2$. In terms of $s$ and $t$,
the diagonal density matrix elements $\varrho_{MM}=\varrho_{-M,-M}$ for the
production of a virtual photon with spin projection $M$ in the
Gottfried--Jackson frame are  \begin{eqnarray}
  \varrho_{11} & = & \frac{2}{(s+t)^2 (2Q^2-s-t)^2}
            \left[ s^5+3s^4t+3s^4Q^2+4s^3t^2+6s^3tQ^2\right.
\nonumber\\
  & &  \mbox{} +4s^2t^3-2s^2t^2Q^2-4s^2Q^6+3st^4-10st^3Q^2+8st^2Q^4+t^5
\nonumber \\
  & & \left. \mbox{} -5t^4Q^2+8t^3Q^4-4t^2Q^6 \right],
\label{ro11}\\
  \varrho_{00} & = & -\frac{8stQ^2}{(s+t)^2}. \label{ro00}
\end{eqnarray}
In the following, we will neglect the target parton transverse momentum $\ktb$
in Eq. (\ref{exact}). In terms of the density matrix elements above, the
polarization parameter $\lambda$ of Eq. (\ref{deflambda}) then becomes:
\begin{equation} \lambda =
\frac{\varrho_{11}-\varrho_{00}}{\varrho_{11}+\varrho_{00}}. \label{lbd}
\end{equation}

In Fig. 2 we show $\lambda$ as a function of $x_F$. Our result
(solid line)
takes the exact kinematics into account.
We compare this
to the BB limit of Eq. (\ref{BBlamb}) (dashed line) and to
the E615  data \cite{Conway}.
We fix $\kta^2 = $ 0.8 GeV$^2$, $ Q =
$ 4.5 GeV and $ p_{\rm beam} = $ 252 GeV
for both calculations, in accordance with the data
\cite{Conway}. There is a sizeable difference between the solid
and dashed curves, indicating the importance of nonasymptotic kinematics at
present energies. Furthermore, the agreement with the data is clearly better
using the exact kinematics. At this value of
$\kta^2$, we find that the general expression (\ref{exact}) gives
values of $\lambda$ consistent within $\Delta\lambda=0.1$ with the
asymptotic limit (\ref{BBexpr}) for $x_F \leq
0.9$ when $Q^2$ and $s$ are scaled by a common factor
$\gsim 25$. Since the mean value of $\kta^2$ should increase with $Q^2$, the
approach to the limit (\ref{BBlamb}) will in reality be even slower.

There are generic reasons for the large finite energy corrections at high
$x_F$. This can be seen, \eg, from the
definition of $t$ in (\ref{t}).
For $\pabs \to \infty$ at fixed $x_a$ and $\kta$,
the leading term of $\order({\bf p^2})$ in
$t$ cancels, giving $t \simeq -\kta^2/(1-x_a)$. On the other hand, for $x_a \to
1$ at fixed (albeit large) $\pabs$ we get the very different result
$t \simeq - 2 \pabs |\kta|$.
A similar sensitivity to the order in which the limits are
taken can be seen in the argument of the first delta function in Eq.
(\ref{exact}). This is the constraint which fixes the relation between the BB
model variables $x_a,\ x_b$ and the measurable quantities $x_F,\ Q^2$. In
Ref. \cite{Conway}, the data was compared to the BB model as a function of
$x_a$, using the relation $x_a = \frac{1}{2}(x_F+\sqrt{x_F^2+Q^2/\pabs^2})$,
which holds only in the asymptotic limit. In Fig. 2, we chose the physical
quantity $x_F$ instead, to avoid ambiguities related to model kinematics.

At present fixed target energies, $\pabs = \order(10 \gev)$, hence
$(1-x_F)\pabs = \order(1 \gev) \simeq \langle |\QT| \rangle$ for $x_F = 0.9$.
Thus it is not surprising to find large deviations due to finite energy
effects. The improved agreement with data obtained in Fig. 2 when such effects
are taken into account is of course encouraging. However, it is also an
indication that one should carefully reconsider the applicability of the twist
expansion of QCD, on which the BB model is based.

In the usual, leading twist QCD approach, the time scale $\tau_I \simeq 2p/Q^2$
of the hard interaction (here $q\bar q \to \mu\mu$) is much shorter than the
lifetime of the Fock state, $\tau_F \simeq 2p(1-x_a)/\kta^2$ (\cf\ Eq.
(\ref{ediff})). In the BB model \cite{BergerBrodsky}, the limit (\ref{BBlimit})
was taken such that $\tau_I/\tau_F = \order(1-x_a)$ is vanishing. Hence the
factorization between the wave function dynamics and that of the hard
scattering subprocess is still valid. On the other hand, in the limit
\cite{BHMT}
\begin{equation}
x_a \to 1, \ \ \  Q^2 \to \infty\ \ {\rm with}\ \ \kta^2 / Q^2  \sim
(1-x_a) \label{bhmt}
\end{equation}
we have $\tau_I \sim \tau_F$ and the dynamics of the subprocess
scattering is inseparable from that of the Fock state. In this limit the twist
expansion breaks down and the two new diagrams shown in Fig. 3 contribute at
leading order to the muon pair production process.

At finite energies, it is not obvious which (if any) of the asymptotic limits
(\ref{BBlimit}), (\ref{bhmt}) is more appropriate. Although one may, as we have
done here, take into account the exact kinematics within a given model, any
application of perturbative QCD must still depend on an idealized high energy
limit. It would be worthwhile to systematically study how the effects of
finite energy corrections may be minimized, since they reflect the inherent
uncertainties of the approach. In the present case, we did check numerically
that including the diagrams of Fig. 3 does not significantly change the
prediction for $\lambda$ given by the solid curve in Fig. 2.

Muon pair production is a good test case for studying large $x_F$ QCD
dynamics, since the number of diagrams is relatively small, and data is
available. This reaction is moreover used to determine the structure
function of the pion, as well as the shadowing effects of antiquarks in nuclei
\cite{E772}. The reliability of these determinations at large $x_F$ depend on
a proper understanding of the dominant reaction mechanism. In the limits
(\ref{BBlimit}), (\ref{bhmt}) the production cross section is not given by the
 projectile structure function but rather by the distribution amplitude, \ie,
the valence wave function at vanishing transverse distance between the partons
\cite{LB}, as indicated in Eq. (\ref{exact}). The pion distribution amplitude
is
in itself of considerable current interest. It is not conclusively settled
whether the exclusive pion form factor data can, at present energies, be
described using the asymptotic hard QCD dynamics of Ref. \cite{LB} or whether
the nonperturbative ``Feynman'' mechanism \cite{ILS} is more appropriate. A
better understanding of large $x_F$ muon pair production could help resolve
this issue, by yielding information on the distribution amplitude.

While completing this work, we learned of a related study of the muon angular
distribution at high $x_F$ \cite{BBKM}. These authors are considering the
effects of terms where the transverse momentum $\QT^2$ of the pair is
comparable
to $Q^2$. In the high energy limit where this ratio is nonvanishing, the
lifetime $\tau_F$ of the Fock state is actually much shorter than the time
scale $\tau_I$ of the hard scattering. This illustrates a further possibility
of
taking limits, not covered by Eqs. (\ref{BBlimit}) and (\ref{bhmt}).

To summarize, we have found that there are general reasons to expect large
finite energy corrections when applying QCD expressions that are valid in
$x_F \to 1$ limits to current data. As a case study, we considered
the model proposed by Berger and Brodsky \cite{BergerBrodsky}, and found that
this model agrees with data on the angular distribution of the muon
pair only when the kinematics are treated exactly.

{\bf Acknowledgement}. We thank S.J. Brodsky for comments on the manu\-script.

\clearpage \begin{center} {\Large \bf Figure Captions} \end{center}

\noindent {\large\bf Fig. 1.}
The relevant Feynman graphs for the Berger and Brodsky
\cite{BergerBrodsky} higher-twist model of Drell-Yan production.

\vskip 0.5 truecm
\noindent {\large\bf Fig. 2.}
The polarization parameter $\lambda$ of Eq. (\ref{deflambda})
as a function of $x_F$. Our calculation (\ref{lbd}) with exact kinematics
is shown by the solid line, the prediction (\ref{BBexpr}) of the
Berger-Brodsky limit (\ref{BBlimit}) by the dashed line, and the data
\cite{Conway} by the boxes. We used $p_{\rm beam}=252$ GeV, $Q=4.5$ GeV,
$\kta^2
= 0.8$ GeV$^2$ and $\ktb^2 = 0$.

\vskip 0.5 truecm
\noindent {\large\bf Fig. 3.}
The two higher-twist diagrams which, together with
the diagrams of Fig. 1, contribute at leading order in the limit (\ref{bhmt}).

\end{document}